# Limits on Dust in Rich Clusters of Galaxies
# From the Color of Background Quasars

Dan Maoz

School of Physics & Astronomy and Wise Observatory
Tel-Aviv University, Tel-Aviv 69978, ISRAEL
dani@wise.tau.ac.il

## Abstract

I measure the $V - I$ color distribution of two samples of radio-selected quasars. Quasars from one sample are projected on the sky within 1° of a rich foreground Abell cluster of galaxies, while quasars from the other sample are more than 3° from any such cluster . There is no significant difference between the color distributions of the two samples. The 90% upper limit on the relative reddening between the two samples is $E(B - V) = 0.05$ mag. This result limits the allowed quantity of smoothly distributed dust in rich clusters, and contradicts previous indications for the existence of such a component.

# 1 Introduction

Distant populations such as faint field galaxies and quasars are often viewed through foreground clusters of galaxies, which can distort our view of those populations. Measuring such distortions can be useful for a better understanding of both the background objects and of the mass-content and structure of the clusters.

Clusters can have two opposite effects on the density of background objects seen through a cluster. Gravitational lensing by the cluster potential can cause either an enhancement or a deficit in the number of background objects, depending on the intrinsic number vs. brightness relation of the background population, and the limiting magnitude to which the population is detected (e.g. Rodrigues-Williams & Hogan 1994). Dust in a cluster will obscure and redden a background population, and will therefore reduce the density of background objects down to a given magnitude limit. The detection of dust in rich clusters is of great interest in itself, since it can provide clues to the source of the observed intracluster X-ray emitting gas (e.g. Dwek, Rephaeli & Mather 1990).

Over the past years there have been various reports of overdensities of quasars behind foregound clusters of galaxies (Bartelmann, Schneider, & Hasinger 1994; Bartelmann & Schneider 1993a; Wu & Han 1995; Seitz & Schneider 1995; Williams & Hawkins 1995). Quasar-galaxy associations have also been reported (e.g. Fugmann 1988, 1990; Bartelmann & Schneider 1993b) and, when occuring on large angular scales, are probably evidence for the same phenomenon (i.e. the effect of the clusters in which the galaxies reside). The magnitude of the effect is still controversial. Some workers find an enhancement that is too large to be consistent with the magnification effect of gravitational lensing, given our current assumptions about the quasar population and the masses of clusters (e.g. Williams & Hawkins 1995).

On the other hand, several studies have reported large deficits of distant quasars or clusters of galaxies behind nearby clusters of galaxies, and have proposed extinction by intracluster dust as the cause. Zwicky (1957) first pointed out a deficit of distant clusters behind rich ones. Karachentsev & Lipovetskii (1969) used the same method to derive a mean cluster extinction of $A_V \approx 0.2$ mag. Similarly, Bogart & Wagoner (1973) found that distant Abell clusters are anticorrelated on the sky with nearby ones. To account for this effect, they argued for an extinction corresponding to $A_V \approx 0.4$ mag and extending to $\sim 2.5$ times the optical radii of the nearby clusters. Szalay, Holloshi & Toth (1989), confirmed the effect but ascribed it to the difficulty of identifying background clusters, rather than to dust. Boyle, Fong, & Shanks (1988) identified quasars and clusters of galaxies on automatically-scanned plates. They found a $\sim 30\%$ deficit of quasars within $4'$ of the clusters, attributable to an extinction of $A_V = 0.15$ mag. Romani & Maoz (1992, hereafter RM92) found that optically-discovered quasars from the Véron-Cetty & Véron (1989) catalog avoid rich foreground Abell clusters of galaxies, with a deficit of $60\%$ within $20' - 40'$ of the cluster, and $\sim 25\%$ out to a radius of $1°$. Radio-selected quasars are free from this effect. RM92 postulated that there is an average extinction of $A_V = 0.4$ within a $1°$ radius of the cluster, but that it is probably patchy, with many of the lines of sight undergoing as much as 1 mag of extinction. Williams and Hawkins (1995) recently searched for a reddening and obscuration effect in variability-selected quasars behind galaxy clusters identified automatically on UK-Schmidt plates. They find an *overdensity* of quasars behind the clusters, which they attribute to gravitational lensing, and no evidence for reddening, based on the comparison of the $U - B$ color distributions of the quasars that are behind clusters and those in the field.

The various results on extinction in clusters quoted above are not necessarily in conflict, because the experiments are not exactly the same. For example, the galaxy clusters defined by Boyle et al. (1988) consist of tens of galaxies, as opposed to the Abell clusters treated by Bogart & Wagoner (1973) and RM92, which are formed of hundreds of galaxies each. As pointed out by RM92, selection effects in the various identification methods of quasars may have also affected some or all of the above results. Several other workers have searched for evidence of dust in clusters through reddening of elliptical galaxies (Ferguson 1993), IR emission (Annis & Jewitt 1993), and emission-line ratios of cooling flows (Hu 1992), but have come up only with upper limits of $E(B - V) \lesssim 0.1 - 0.2$ mag. An intercomparison of various results on dust in clusters is given by Williams (1995).

In this paper, I search for a signature of dust in rich clusters and attempt to disentangle it from selection effects by testing the following prediction of RM92. If intracluster dust is responsible for the avoidance of rich Abell clusters by optically-selected background quasars, radio-selected background quasars projected within 1°
of Abell clusters will be redder than those that are not.

The $V-I$ color of quasars is a particularly accessible probe of foreground reddening. The relatively small scatter in the power-law slope of quasars, $\alpha \approx -0.5 \pm 0.4$ (e.g. Richstone & Schmidt 1980), translates into a narrow $V-I$ distribution, with $V-I = 0.65 \pm 0.25$. The Ly$\alpha$ forest, with its effect on the spectral slope of quasars, enters the $V$ band only beyond redshift $z \sim 3$, and hence does not affect the $V-I$ color of most known radio quasars. Because of the large wavelength interval between the $V$ and $I$ bands, $V-I$ is a sensitive reddening indicator. Foreground reddening will shift the mean of the $V-I$ distribution of quasars, relative to an unreddened quasar sample. The accuracy to which the mean is measured is limited only by the number of quasars observed.

## 2  Sample Selection, Observations, and Analysis

To test for the presence of reddening in clusters I defined two samples of quasars. Quasars in one sample (hereafter "the obscured sample") are projected on the sky within 1° of at least one rich and distant ($R \geq 1, D \geq 5$) Abell cluster of galaxies. (See Abell 1958, for definition of richness and distance classes, $R$, and $D$.) These criteria correspond to the population and the angle of the avoidance effect found by RM92. The second sample (hereafter "the unobscured sample") consists of quasars that are at least 3° from any such cluster. Great care was exercised in choosing the quasar samples and observing them, in order to avoid artificially introducing color differences between the samples through selection effects or systematic errors, but also to minimize any noise that may dilute an existing signal.

From Véron-Cetty & Véron (1993) I selected all quasars from the PKS, B2, 3C, and 4C radio surveys with redshift $z > 0.3$, positive declination, Galactic latitude $|b^{II}| > 40°$ for longitude $60° < l^{II} < 300°$ and $|b^{II}| > 50°$ otherwise, and apparent magnitude $V \leq 18.9$. From among these quasars, I then selected all that were either within 1° of at least one $D \geq 5$, $R \geq 1$ cluster of galaxies listed by Abell (1958), or more than 3° from any such Abell cluster.

The motivation for choosing radio-selected quasars was that such quasars were shown by RM92 to *not* avoid clusters, and should therefore be unbiased probes of the reddening induced by the clusters. (For example, if dust extinction causes the avoidance of clusters by optically-selected quasars, then optically-selected quasars near clusters would preferably be those that had escaped extinction, and hence would be relatively unreddened.) Only quasars from the four radio surveys above were chosen, in order to make the sample as homogeneous as possible while not making it too small. The redshift limit on the quasars ensured that they are behind the clusters; Struble & Rood (1991) have shown that the redshifts of the $D \geq 5$ clusters are $0.1 \lesssim z \lesssim 0.3$. Positive declinations were chosen to allow measurements at low airmass, where the error in color determination is smallest. The limits on Galactic coordinates avoided regions of the sky where the Abell catalog is substantially incomplete due to Galactic extinction. This is important especially in choosing the "unobscured" sample, since quasars behind uncatalogued clusters would be mixed into this sample, and hence potentially dilute a difference between the obscured and unobscured samples. The high Galactic latitude of all the quasars also guarantees that the additional scatter induced by Galactic reddening is at most $\pm 0.02$ mag (Burstein & Heiles 1978). The apparent-magnitude limit was chosen to allow accurate measurement of a large sample in a short time. The 1° limit corresponds to the radius of the effect found by RM92. The 3° radius is the largest that still produces a quasar sample large enough to statistically detect a small ($\gtrsim 0.1$ mag) shift in the color distribution.

Application of the above criteria produced a list of 41 quasars in the obscured sample and 46 quasars in the unobscured sample. Tables 1 and 2 list the names and parameters of the selected quasars. Figure 1 shows the distribution in Galactic coordinates of the quasars and the Abell clusters. Note that the two quasar samples are well mixed on the sky. They are also similar in their redshift and magnitude distributions.

The quasars were imaged with the Wise Observatory 1m telescope on 1995 May 3-4, May 8-9, and July 28-29 UT using a Tektronix $1024 \times 1024$-pixel back-illuminated CCD at the Cassegrain focus. Standard Johnson-Cousins $V$ and $I$ filters were used. Photometric standards from Landolt (1992) were observed throughout each night. Exposure times were 300 s in each filter for quasars with estimated catalog magnitudes fainter than 17.9 mag, and 150 s in each filter for the brighter ones. To avoid systematic effects, the observations of the obscured and unobscured samples were intermingled as much as possible on each night and between nights.

Standard CCD data reduction was carried out with the IRAF package. The instrumental $V$ and $I$ magnitudes of each quasar were measured with the DAOPHOT (Stetson 1987) task within IRAF, relative to bright stars in the same image that served to model the point-spread function (PSF). Typical uncertainties in the PSF fitting were 0.01 mag for the brighter quasars to 0.04 mag for the fainter ones. Aperture photometry of the standard stars was used to calculate a photometric solution and its errors, which was applied to the quasar measurements to bring them to the Landolt (1992) system. Color-correction terms were small ($\sim 0.01$ mag). The scatter of the standard-star measurements about the photometric solution was 0.02–0.04 mag. This error was combined in quadrature with the DAOPHOT error and with the photometric error as determined from the covariance matrix of the photometric solution, to form the total error on the $V$ and $I$ magnitude. The $V$ and $I$ measurements, the $V - I$ color, and their errors appear in Tables 1 and 2. Typical errors in $V - I$ are 0.07 mag. Due to observing contraints (weather, moonlight, time), 31 quasars from the obscured sample and 40 quasars from the unobscured sample were successfully observed. To reduce scatter, five quasars from the obscured sample and five quasars from the unobscured sample with $V - I$ errors exceeding 0.1 mag will be excluded from the subsequent analysis.

Figure 2 shows the binned $V - I$ distributions of the obscured and unobscured samples. There is no obvious difference between the two samples. A Kolmogorov-Smirnoff (KS) test on the cumulative distributions fails to reject the null hypothesis that the two samples are drawn from the same population. The mean and the standard deviation of the mean is $0.655 \pm 0.043$ and $0.623 \pm 0.048$ for the unobscured and obscured samples, respectively. A Student's t-test shows that the two means are consistent. From Student's-t distribution, the 95% (90%) upper limit on the difference between the means of the two samples is $\Delta(V - I) = 0.108$ mag (0.084 mag). There is also no evidence for a bimodality in the distribution of the obscured sample, as would be expected if there was patchy obscuration in the clusters intercepting of order half the lines of sight. There are several abnormally red ($V - I > 1$) quasars, but these exist in equal proportions in both the obscured and the unobscured samples.

To test for systematic effects in the calibration, the $V - I$ distributions of various nights were compared. There are no significant differences based on a KS test or a Student's t-test.

## 3 Discussion and Conclusions

For Galactic dust, $E(V - I) = 1.60E(B - V) = 0.52A_V$ (Rieke & Lebofski 1985). Assuming such properties for intracluster dust (see Dwek et al. 1990, for a discussion of the suitability of such an assumption), the upper limit on the reddening found above, $E(V - I) < 0.084$ mag, translates to $E(B - V) < 0.053$ mag or $A_V < 0.162$ mag at the 90% confidence level.

RM92 found suggestions in the limited color data for the quasars in the Véron catalog that, if dust is responsible for the avoidance of clusters by quasars, the obscuration is patchy, with about half of the lines of sight suffering a reddening of $E(B - V) \approx 0.3$ mag, or $E(V - I) \approx 0.5$. Such an effect would have been readily apparent in the data presented here, and is strongly ruled out. Similarly, the average $A_B \approx 0.5$ mag (i.e. $A_V \approx 0.4$mag) predicted by RM92 is ruled out. Since most of the quasars measured here are $> 30'$ from the cluster centers, a centrally-peaked dust distribution, with high central reddening and $A_V \approx 0.11$ mag out to $1°$, cannot be ruled out. Dwek et al. (1992) have, however, argued that dust can only survive in the outer parts of the cluster, if at all. Note also that the seven quasars that are $< 30'$ from the clusters do not have unusual color. There is no trend of quasar color with angular separation from the clusters or with angular separation normalized by the Leir & van den Bergh (1977) angular radius of each cluster.

I conclude that, as alternatively proposed by RM92, the avoidance of foreground clusters by optically-discovered quasars is probably a selection effect due to the difficulty of identifying quasars (whether using objective-prism or color excess techniques) in crowded fields. A similar, selection-induced avoidance of Galactic stars by quasars found in objective-prism surveys has been demonstrated by Gould, Bahcall, & Maoz (1993).

Dust in rich clusters might still be accomodated with the present results if both the comoving density *and* the dust content of the clusters do not evolve between $z = 0$ and the mean redshift of the quasars, $z = 1.22$. Under this assumption, the line of sight to the average quasar in either sample passes through the postulated $\sim 6$-Mpc dust halo of $\sim 2 - 3$ distant clusters, and both samples would be equally reddened. If this were the case, however, one would expect to see at $0.3 < z \lesssim 1.2$ a correlation of quasar $V - I$ color with redshift. I find no such trend in the data, but the number of objects is small.

The results presented here also argue against the extinctions of $A_V \approx 0.2$ mag found by Karachentsev & Lipovetskii (1969) and $A_V \approx 0.4$ mag derived by Bogart & Wagoner (1973) for *nearby* Abell clusters, based on their anticorrelation with distant clusters. The difficulty in identiying several clusters on one line of sight is the likely true cause of the effect. This is not necessarily the case, however, if the dust properties of Abell clusters are strongly evolving, such that dust has appeared in the intracluster medium of nearby clusters by $z \lesssim 0.1$. The results I have shown only marginally rule out the small amount of extinction ($A_V \approx 0.15$) deduced by Boyle et al. (1988), and in any case, that effect was reported for a different (much poorer) population of clusters. That result, however, is challenged by the null detection of reddening for a similar population of clusters by Williams & Hawkins (1995). They propose that Boyle et al.'s (1988) reported underdensity of quasars behind clusters is the result of gravitational lensing by the cluster potential. Such an effect can arise as a result of a faint quasar-detection flux limit combined with a shallow-sloped quasar number-brightness relation.

To summarize, I have shown that radio-selected quasars behind Abell clusters are not redder than quasars that do not have a cluster in the foreground. The 95% upper limit on the reddening exerted by the intracluster medium is $E(V - I) < 0.108$ mag. This result suggests that the avoidance of foreground Abell clusters by optically-selected quasars is, at least in large part, a selection effect rather than a signature of dust extinction. The limit on reddening presented here improves upon previous limits based on null detections of reddening effects, and constrains the mean column density of dust in a rich cluster to $< 6.5 \times 10^{-6}$ g cm$^{-2}$, and the total dust mass within a 1° radius to $\lesssim 1.5 \times 10^{12} M_\odot$. Smaller amounts of reddening in clusters, as suggested by some studies, are still allowed. The bounds on dust in clusters can be improved by measuring the color distributions of larger samples of quasars.

**Acknowledgements** I thank O. Bergman, J. Dan, and N. and M. Gardosh, for their invaluable assistance with the observations, and N. Brosch, R. Romani, A. Sternberg, and L.R. Williams, for useful comments and suggestions. Astronomy at Wise Observatory is supported by grants from the Israel Academy of Sciences and the Ministry of Science and Technology.

# References

Abell, G.O. 1958, ApJS, 3, 211

Annis, J. & Jewitt, D. 1993, MNRAS 264, 593

Bartelmann, M., Schneider, P., & Hasinger, G. 1994, A&A, 290, 399

Bartelmann, M., & Schneider, P. 1993a, A&A, 268, 1

Bartelmann, M., & Schneider, P. 1993b, A&A, 271, 421

Bogart, R.S. & Wagoner, R.V. 1973, ApJ, 181, 609

Boyle, B.J., Fong, R. & Shanks, T. 1988, MNRAS, 231, 897

Burstein, D., & Heiles, C. 1978, ApJ, 225, 40

Dwek, E., Rephaeli, Y. & Mather, J.C. 1990, Ap J, 350, 104

Ferguson, H.C. 1993, MNRAS, 263, 343

Fugmann, W. 1988, A&A, 204, 73

Fugmann, W. 1990, A&A, 240, 11

Gould, A., Bahcall, J.N., and Maoz, D. 1993, ApJS, 88, 53

Hu, E.M. 1992, ApJ, 391, 608

Karachentsev, I.D., & Lipovetskii, V.A. 1969, Soviet Astron., 12, 909

Landolt, A.U. 1992, AJ, 104, 340

Leir, A.A., & van den Bergh, S. 1977, ApJS, 34, 381

Richstone, D.O., & Schmidt, M. 1980, ApJ, 235, 361

Rieke, G.H., & Lebofsky, M.J. 1985, ApJ, 288, 618

Rodrigues-Williams, L.L., & Hogan, C.J. 1994, AJ, 107, 451

Romani, R.W., & Maoz, D. 1992, ApJ, 386, 36 (RM92)

Seitz, S., & Schneider, P. 1995, A&A, in press

Stetson, P.B. 1987, PASP, 99, 191

Struble, M., & Rood, H. 1991, ApJS, 77, 363

Szalay, A.S., Hollósi, J. & Tóth, G. 1989, ApJ, 339, L5

Véron-Cetty, M.-P., & Véron, P. 1989, "A Catalogue of Quasars and Active Nuclei" (4th Edition) (Munich: ESO)

Véron-Cetty, M.-P., & Véron, P. 1993, "A Catalogue of Quasars and Active Nuclei" (6th Edition) (Munich:ESO)

Williams, L.L.R., & Hawkins, M.R.S. 1995, in preparation

Williams, L.L.R. 1995, Ph.D. thesis, University of Washington

Wu, X.-P., & Han, J. 1995, MNRAS 272, 705

Zwicky, F. 1957, "Morphological Astronomy" (Berlin: Springer Verlag), 101

Table 1: Quasars $> 3°$ from Abell clusters

| Name | R.A.(1950) | Dec.(1950) | z | V | $\delta V$ | I | $\delta I$ | $V - I$ | $\delta(V - I)$ |
|---|---|---|---|---|---|---|---|---|---|
| PKS 0109+17 | 01:09:09.6 | $17°37'56"$ | 2.157 | 18.940 | 0.059 | 18.083 | 0.064 | 0.857 | 0.087 |
| PKS 0114+07 | 01:14:49.5 | $07°26'30"$ | 0.861 | 18.374 | 0.053 | 18.091 | 0.061 | 0.284 | 0.081 |
| PKS 0229+13 | 02:29:02.4 | $13°09'40"$ | 2.065 | − | − | − | − | − | − |
| PKS 0256+075 | 02:56:47.0 | $07°35'45"$ | 0.893 | − | − | − | − | − | − |
| B2 0901+28B | 09:01:30.6 | $28°31'31"$ | 1.121 | 17.829 | 0.051 | 17.211 | 0.052 | 0.618 | 0.072 |
| B2 0941+26 | 09:41:50.2 | $26°08'33"$ | 2.906 | 18.591 | 0.063 | 17.918 | 0.057 | 0.673 | 0.085 |
| 4C 40.24 | 09:45:50.0 | $40°53'43"$ | 1.252 | − | − | − | − | − | − |
| PKS 0952+179 | 09:52:11.8 | $17°57'44"$ | 1.472 | − | − | − | − | − | − |
| 3C 232 | 09:55:25.5 | $32°38'23"$ | 0.533 | 16.546 | 0.039 | 16.306 | 0.040 | 0.240 | 0.056 |
| PKS 1012+022 | 10:12:40.8 | $02°13'50"$ | 1.374 | 17.493 | 0.047 | 16.666 | 0.045 | 0.828 | 0.065 |
| 4C 60.15 | 10:45:23.3 | $60°24'37"$ | 1.722 | 18.862 | 0.063 | 17.731 | 0.049 | 1.130 | 0.080 |
| PKS 1048+24 | 10:48:46.7 | $24°03'58"$ | 1.270 | − | − | − | − | − | − |
| B2 1104+36 | 11:04:41.6 | $36°32'26"$ | 0.393 | 18.942 | 0.066 | 18.226 | 0.069 | 0.716 | 0.096 |
| 3C 261 | 11:32:16.3 | $30°22'01"$ | 0.614 | 18.562 | 0.051 | 17.686 | 0.046 | 0.876 | 0.068 |
| PKS 1134+01 | 11:34:55.7 | $01°32'51"$ | 0.430 | 18.445 | 0.115 | 17.245 | 0.057 | 1.200 | 0.128 |
| B2 1148+38 | 11:48:53.3 | $38°42'33"$ | 1.299 | 17.227 | 0.040 | 16.544 | 0.040 | 0.683 | 0.057 |
| B2 1204+39 | 12:04:04.6 | $39°57'15"$ | 1.530 | 18.357 | 0.051 | 17.726 | 0.055 | 0.631 | 0.076 |
| 3C 268.4 | 12:06:42.1 | $43°56'02"$ | 1.400 | 18.109 | 0.047 | 16.934 | 0.038 | 1.175 | 0.061 |
| B2 1211+33 | 12:11:32.8 | $33°26'26"$ | 1.598 | 17.551 | 0.041 | 16.749 | 0.042 | 0.802 | 0.058 |
| 4C 53.24 | 12:13:01.5 | $53°52'36"$ | 1.065 | 18.245 | 0.043 | 17.799 | 0.061 | 0.446 | 0.075 |
| 3C 270.1 | 12:18:03.9 | $33°59'50"$ | 1.519 | 18.520 | 0.048 | 17.855 | 0.063 | 0.664 | 0.079 |
| B2 1220+37 | 12:20:42.3 | $37°23'39"$ | 0.489 | 18.346 | 0.042 | 17.756 | 0.074 | 0.590 | 0.085 |
| B2 1225+31 | 12:25:55.9 | $31°45'13"$ | 2.219 | 15.866 | 0.038 | 15.147 | 0.036 | 0.719 | 0.052 |
| B2 1234+33B | 12:34:36.8 | $33°30'53"$ | 1.280 | 18.426 | 0.048 | 18.276 | 0.093 | 0.150 | 0.105 |
| PKS 1236+077 | 12:36:52.3 | $07°46'45"$ | 0.400 | 19.180 | 0.057 | 17.810 | 0.046 | 1.370 | 0.073 |
| B2 1237+35 | 12:37:55.4 | $35°19'26"$ | 1.194 | 17.363 | 0.036 | 16.943 | 0.050 | 0.420 | 0.062 |
| 4C 44.20 | 12:39:57.0 | $44°12'34"$ | 0.610 | 18.289 | 0.048 | 17.679 | 0.063 | 0.609 | 0.079 |
| B2 1244+32B | 12:44:55.4 | $32°25'23"$ | 0.949 | 17.509 | 0.041 | 17.151 | 0.042 | 0.358 | 0.059 |
| PKS 1252+11 | 12:52:07.7 | $11°57'21"$ | 0.870 | 16.802 | 0.032 | 16.512 | 0.044 | 0.290 | 0.054 |
| B2 1256+39 | 12:56:41.9 | $39°16'23"$ | 0.978 | 19.459 | 0.085 | 18.577 | 0.077 | 0.882 | 0.115 |
| B2 1300+39 | 13:00:29.0 | $39°46'08"$ | 2.436 | 18.639 | 0.061 | 18.261 | 0.110 | 0.378 | 0.126 |
| 3C 281 | 13:05:22.5 | $06°58'14"$ | 0.599 | 17.347 | 0.041 | 16.866 | 0.042 | 0.481 | 0.059 |
| B2 1306+27A | 13:06:33.1 | $27°24'10"$ | 1.537 | 18.337 | 0.043 | 17.670 | 0.060 | 0.667 | 0.073 |
| 4C 18.36 | 13:08:29.5 | $18°15'34"$ | 1.689 | 19.004 | 0.071 | 18.216 | 0.061 | 0.788 | 0.093 |
| 3C 287.0 | 13:28:15.9 | $25°24'38"$ | 1.055 | 17.985 | 0.039 | 17.648 | 0.071 | 0.338 | 0.081 |
| 3C 286.0 | 13:28:49.7 | $30°45'58"$ | 0.846 | 17.271 | 0.040 | 16.773 | 0.038 | 0.497 | 0.056 |
| PKS 1335+023 | 13:35:06.9 | $02°22'12"$ | 1.356 | 18.386 | 0.049 | 17.687 | 0.086 | 0.699 | 0.099 |
| B2 1409+34A | 14:09:46.2 | $34°29'16"$ | 1.820 | 18.457 | 0.045 | 17.740 | 0.044 | 0.716 | 0.063 |
| B2 1414+34 | 14:14:50.0 | $34°42'40"$ | 0.750 | 17.586 | 0.032 | 17.426 | 0.052 | 0.161 | 0.061 |
| 4C 46.29 | 14:15:13.5 | $46°20'55"$ | 1.552 | 18.150 | 0.043 | 17.417 | 0.045 | 0.733 | 0.062 |
| PKS 1421+122 | 14:21:04.6 | $12°13'26"$ | 1.611 | 18.377 | 0.038 | 17.720 | 0.064 | 0.657 | 0.074 |
| PKS 1427+109 | 14:27:43.8 | $10°56'46"$ | 1.710 | 18.872 | 0.045 | 18.446 | 0.111 | 0.426 | 0.120 |
| B2 1452+30 | 14:52:25.0 | $30°08'06"$ | 0.580 | 19.028 | 0.053 | 18.268 | 0.056 | 0.759 | 0.078 |
| B2 1612+37A | 16:12:03.7 | $37°50'29"$ | 1.630 | 18.896 | 0.043 | − | − | − | − |
| B2 1633+38 | 16:33:30.6 | $38°14'09"$ | 1.814 | 17.665 | 0.033 | 16.845 | 0.042 | 0.820 | 0.053 |
| PKS 2301+060 | 23:01:56.3 | $06°03'57"$ | 1.268 | 18.433 | 0.045 | 17.810 | 0.060 | 0.623 | 0.075 |

Table 2: Quasars < 1° from Abell clusters

| Name | R.A.(1950) | Dec.(1950) | z | V | δV | I | δI | V − I | δ(V − I) | Sep.['] (Abell No.) |
|---|---|---|---|---|---|---|---|---|---|---|
| PKS 0003+15 | 00:03:25.1 | 15°53'07" | 0.450 | 15.614 | 0.035 | 15.270 | 0.045 | 0.343 | 0.057 | 43.1(0001), 23.6(2705) |
| PKS 0008+171 | 00:07:59.4 | 17°07'38" | 1.601 | 17.333 | 0.037 | 17.018 | 0.052 | 0.315 | 0.064 | 17.1(0006) |
| 4C 08.04 | 00:33:41.0 | 07°58'34" | 1.578 | 19.081 | 0.078 | 18.291 | 0.076 | 0.790 | 0.109 | 56.0(0068) |
| 4C 09.01 | 00:33:48.3 | 09°51'29" | 1.909 | 18.310 | 0.052 | 17.254 | 0.055 | 1.056 | 0.076 | 59.6(0073) |
| 3C 37 | 01:15:43.6 | 02°42'20" | 0.670 | 18.817 | 0.058 | 17.986 | 0.053 | 0.832 | 0.079 | 45.3(0172) |
| 3C 47.0 | 01:33:40.5 | 20°42'09" | 0.425 | 17.905 | 0.040 | 17.339 | 0.049 | 0.567 | 0.063 | 27.7(0213) |
| PKS 0136+176 | 01:36:59.3 | 17°37'56" | 2.716 | 19.019 | 0.060 | 18.447 | 0.079 | 0.572 | 0.099 | 34.7(0221), 17.6(0227) |
| PKS 0158+18 | 01:58:56.0 | 18°22'09" | 0.799 | 17.714 | 0.039 | 17.393 | 0.049 | 0.321 | 0.062 | 14.7(0288) |
| PKS 0159+034 | 01:59:14.7 | 03°29'03" | 1.897 | 17.221 | 0.039 | 16.519 | 0.051 | 0.702 | 0.064 | 04.8(0293) |
| PKS 0214+10 | 02:14:26.7 | 10°50'18" | 0.408 | 16.456 | 0.037 | 15.993 | 0.047 | 0.463 | 0.060 | 43.9(0331) |
| 4C 47.29 | 08:59:40.0 | 47°02'57" | 1.462 | − | − | − | − | − | − | 35.7(0739) |
| 3C 216.0 | 09:06:17.3 | 43°05'58" | 0.668 | − | − | − | − | − | − | 44.0(0758) |
| PKS 0922+14 | 09:22:22.4 | 14°57'23" | 0.896 | − | − | − | − | − | − | 46.0(0795) |
| B2 0949+36 | 09:49:26.6 | 36°20'12" | 2.050 | 18.359 | 0.048 | 17.693 | 0.059 | 0.666 | 0.076 | 42.9(0893) |
| B2 0952+35 | 09:52:49.3 | 35°47'38" | 1.241 | 19.194 | 0.112 | 18.412 | 0.110 | 0.781 | 0.157 | 42.6(0893) |
| 4C 55.17 | 09:54:14.3 | 55°37'17" | 0.901 | − | − | − | − | − | − | 12.0(0899) |
| B2 1009+33 | 10:09:17.6 | 33°24'17" | 2.260 | − | − | − | − | − | − | 46.8(0943) |
| B2 1011+25 | 10:11:05.6 | 25°04'10" | 1.636 | − | − | − | − | − | − | 42.4(0964) |
| B2 1018+34 | 10:18:24.1 | 34°52'31" | 1.400 | − | − | − | − | − | − | 04.7(0982) |
| PKS 1020+191 | 10:20:11.9 | 19°08'46" | 2.136 | − | − | − | − | − | − | 05.0(0991), 43.2(0994) |
| B2 1020+40 | 10:20:14.6 | 40°03'27" | 1.254 | 17.882 | 0.049 | 17.236 | 0.041 | 0.646 | 0.064 | 52.1(0972) |
| 4C 19.34 | 10:22:01.4 | 19°27'34" | 0.828 | − | − | − | − | − | − | 38.6(0991), 36.5(0994) |
| 4C 09.37 | 10:47:48.9 | 09°41'48" | 0.786 | − | − | − | − | − | − | 59.4(1093), 25.5(1105), 50.2(1115) |
| 3C 254.0 | 11:11:53.2 | 40°53'41" | 0.734 | 17.593 | 0.041 | 17.147 | 0.063 | 0.445 | 0.075 | 37.9(1190), 51.6(1203) |
| PKS 1158+007 | 11:58:49.5 | 00°45'10" | 1.383 | 18.825 | 0.087 | 18.558 | 0.150 | 0.267 | 0.173 | 53.8(1445) |
| B2 1204+28 | 12:04:55.0 | 28°11'42" | 2.177 | 18.426 | 0.054 | 17.803 | 0.046 | 0.622 | 0.071 | 52.0(1455) |
| B2 1208+32A | 12:08:05.6 | 32°13'48" | 0.388 | 16.806 | 0.039 | 16.318 | 0.038 | 0.487 | 0.055 | 55.1(1498) |
| B2 1343+38 | 13:43:26.5 | 38°38'12" | 1.844 | 17.847 | 0.041 | 16.924 | 0.038 | 0.923 | 0.056 | 21.1(1785) |
| B2 1353+30 | 13:53:26.1 | 30°38'51" | 1.018 | 18.365 | 0.041 | 17.498 | 0.074 | 0.867 | 0.085 | 39.5(1826) |
| B2 1425+26 | 14:25:21.9 | 26°45'38" | 0.366 | 16.412 | 0.030 | 16.112 | 0.044 | 0.300 | 0.054 | 36.5(1912) |
| B2 1426+29 | 14:26:32.6 | 29°32'26" | 1.421 | 18.644 | 0.048 | 18.120 | 0.055 | 0.524 | 0.073 | 55.2(1929) |
| 3C 309.1 | 14:58:56.6 | 71°52'11" | 0.904 | 17.269 | 0.031 | 17.113 | 0.100 | 0.156 | 0.105 | 41.8(2037) |
| PKS 1502+106 | 15:02:00.1 | 10°41'18" | 1.833 | 19.164 | 0.063 | 18.087 | 0.069 | 1.077 | 0.093 | 56.9(2016) |
| PKS 1502+036 | 15:02:35.7 | 03°38'08" | 0.413 | 18.709 | 0.048 | 18.000 | 0.078 | 0.709 | 0.091 | 45.9(2023) |
| B2 1506+33A | 15:06:22.6 | 33°58'25" | 2.200 | 18.744 | 0.036 | 17.546 | 0.063 | 1.198 | 0.072 | 40.5(2034) |
| B2 1512+37 | 15:12:46.8 | 37°01'55" | 0.370 | 16.447 | 0.039 | 16.095 | 0.038 | 0.352 | 0.054 | 49.4(2042) |
| 4C 43.39 | 16:29:38.6 | 43°55'03" | 1.167 | 18.779 | 0.050 | 18.392 | 0.055 | 0.387 | 0.074 | 54.3(2190), 36.0(2198), 42.5(2206) |
| 4C 58.32 | 16:34:19.8 | 58°54'42" | 0.985 | 18.950 | 0.058 | 18.472 | 0.067 | 0.478 | 0.088 | 48.2(2208) |
| 4C 47.44 | 16:36:19.2 | 47°23'29" | 0.740 | 18.727 | 0.040 | 18.016 | 0.074 | 0.711 | 0.084 | 55.3(2219) |
| PKS 2318+02 | 23:18:14.4 | 02°40'34" | 1.968 | 18.888 | 0.067 | 18.506 | 0.091 | 0.382 | 0.113 | 53.5(2574), 21.7(2582) |
| PKS 2319+07 | 23:20:03.9 | 07°55'33" | 2.090 | 18.431 | 0.050 | 17.793 | 0.055 | 0.638 | 0.074 | 39.5(2594) |

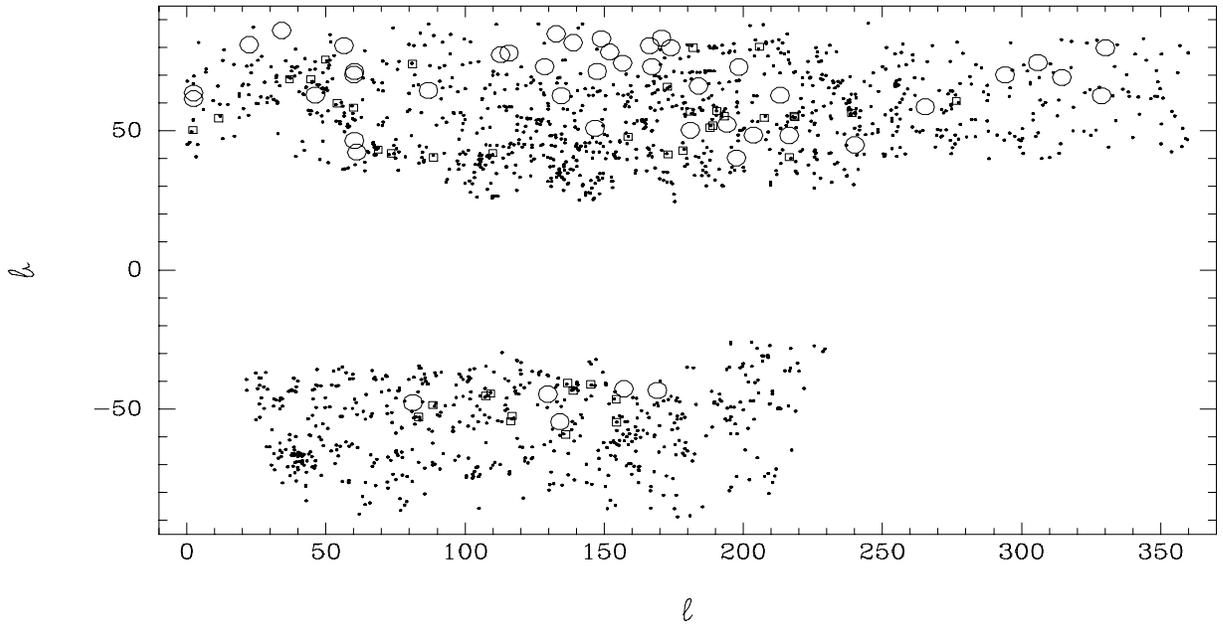

Figure 1: Distribution in Galactic coordinates of quasars and Abell clusters. Points denote all $R \geq 1$, $D \geq 5$ Abell clusters. Squares denote radio-quasars from the "obscured" sample and are $< 1°$ from a foreground cluster. The circles, of radius $3°$, mark the quasars from the "unobscured" sample, which are at least that angular distance from any cluster.

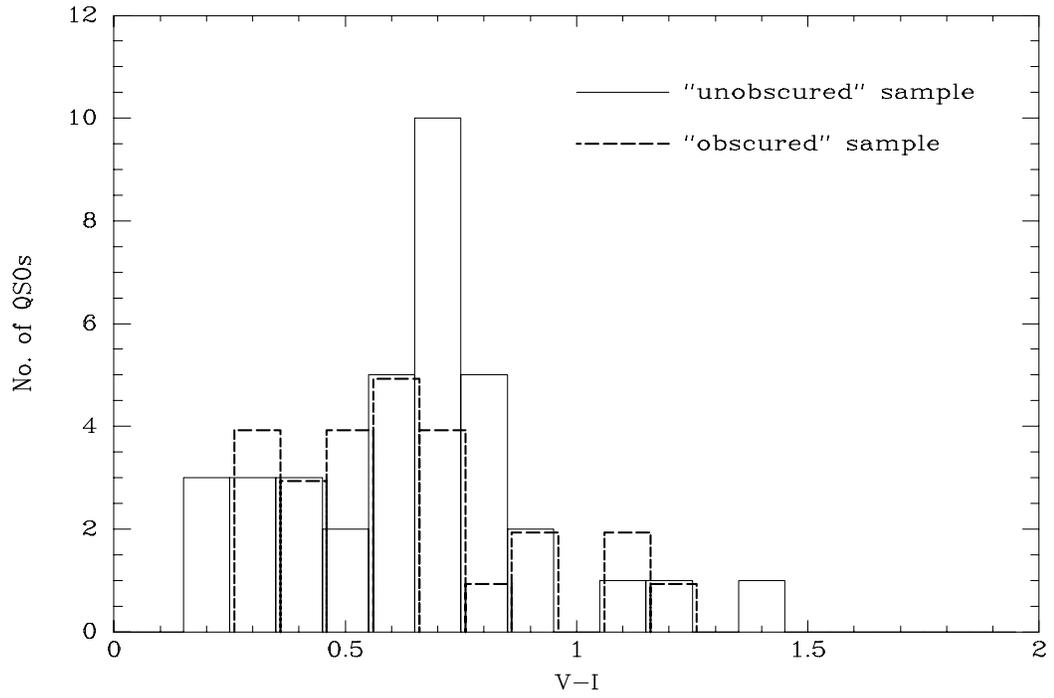

Figure 2: The observed distribution in $V - I$ of the unobscured sample (solid line) and the obscured sample (dashed line) of quasars. There is no significant difference in the distributions or their means, setting a 95% upper limit of $E(V - I) = 0.108$ on the reddening.